# Design of a Monopole Antenna Based Resonant Nanocavity for Detection of Optical Power from Hybrid Plasmonic Waveguides

Kelvin J. A. Ooi,[1,2] Ping Bai,[1,*] Ming Xia Gu,[1] and Lay Kee Ang[2]

[1]Plasmonics and Nanointegration Group, A*STAR Institute of High Performance Computing, 1 Fusionopolis Way, #16-16 Connexis, Singapore 138632

[2]School of Electrical and Electronic Engineering, Nanyang Technological University, Block S1, 50 Nanyang Avenue, Singapore 639798

*Corresponding author: baiping@ihpc.a-star.edu.sg

**Abstract:** A novel plasmonic waveguide-coupled nanocavity with a monopole antenna is proposed to localize the optical power from a hybrid plasmonic waveguide and subsequently convert it into electrical current. The nanocavity is designed as a Fabry-Pérot waveguide resonator, while the monopole antenna is made of a metallic nanorod directly mounted onto the metallic part of the waveguide terminal which acts as the conducting ground. The nanocavity coincides with the antenna feed sandwiched in between the antenna and the ground. Maximum power from the waveguide can be coupled into, and absorbed in the nanocavity by means of the field resonance in the antenna as well as in the nanocavity. Simulation results show that 42% optical power from the waveguide can be absorbed in a germanium filled nanocavity with a nanoscale volume of $220\times150\times60\text{nm}^3$. The design may find applications in nanoscale photo-detection, subwavelength light focusing and manipulating, as well as sensing.

## 1. Introduction

Plasmonics, which allows manipulation of electromagnetic waves at subwavelength scale by using metallic nanostructures, shows great potential for realization of the integration of optics and electronics on the same chip [1]. There are tremendous efforts in developing plasmonic devices that generate [2 – 4], guide [5 – 10], detect [11 – 14], and modulate [15 – 17] electromagnetic waves far below the diffraction limit. Plasmonic waveguides that overcome the overheating and bandwidth limitations of the classical copper interconnects have been considered as potential candidates for future nanophotonic-electronic integrated circuits [1]. Much of the progress has been made on developing low loss, highly confined, and long propagating plasmonic waveguides to replace the classical copper interconnects, among which are the hybrid plasmonic waveguides that demonstrate deep bilateral subwavelength confinement with low propagation loss [5 – 10]. Next, it is essential to develop nanoscale photodetectors that allow for monolithic integration of the plasmonic waveguides with the electronic circuits. However, the traditional photodetectors require very large active materials to ensure near full absorption of the input energy [18, 19]. Linearly scaling down the photodetector to match the size of the subwavelength plasmonic waveguide would dramatically reduce the quantum efficiency of the detector. Therefore, plasmonic nanocavity with antenna has to be used as the basic structure for plasmonic photodetectors, in which plasmonic waves from waveguides are resonantly coupled into a nanocavity filled with active material, to be absorbed and converted into electric signals for further information processing. Our recent work that introduces a dipole nanoantenna to couple the propagating plasmonic waves to a subwavelength nanocavity and to enhance the fields in the nanocavity via localized surface plasmon effect [14] provides a way to design the required photodetector.

The dipole antennas that are used for waveguide-coupled-nanocavities include the open-sleeve dipole [11], straight-rod dipole [13] and L-shaped dipole [14]. A significant amount of the electromagnetic energy is coupled into the nanocavity at the resonance condition of the dipole nanoantenna [13]; however, we found that power confinement and absorption remain low because the antenna feed is not designed as a resonant cavity. On the other hand, nanorod extrusions perpendicular to the dipole along the antenna feed (which results in the antennas looking like L-shapes) serve as an extension for the cavity length to support the longitudinal cavity modes [14], and as a result the magnitudes of the power confinement and absorption are increased by 4 times. This shows that the antenna and nanocavity are correlated for effective reception and localization of the electromagnetic power.

In our previous designs [13, 14], for electrical isolation purposes, the nanocavity has to be separated from the waveguide terminal by a gap. The gap is required to support the formation of the dipole antenna arms. However, the disadvantage of having a gap is that the coupling efficiency from the waveguide into the nanocavity is reduced due to the gap leakage. The successful invention of hybrid plasmonic waveguides [8 – 10] offers innovation possibilities in antenna designs. In this paper, we propose a novel nanocavity design with a monopole antenna, by taking advantage of the hybrid waveguide structure.

**2. Principles for cavity design**

Here, we investigate a structure comprising of a hybrid plasmonic waveguide coupled to a monopole antenna and nanocavity as shown in Fig. 1. The hybrid plasmonic waveguide is formed by sandwiching a thin layer of low-index dielectric (silica) between a metal (aluminum) slab and high-index dielectric (silicon) nanowire. The width and the thickness of the dielectric slot are chosen to be 150nm and 220nm respectively, which takes device fabrication issues into consideration, and also to ensure that the supported plasmonic modes are highly confined in the dielectric slot. The monopole antenna is mounted at the end of the waveguide.

In antenna theory, the monopole is a class of antenna which consists of a straight rod conductor mounted perpendicularly on a large, conductive surface referred to as the ground plane. The monopole is in essence a half-dipole: it can be visualized as being formed by replacing one half of a dipole antenna with a ground plane at right-angle to the remaining half. If the ground plane is large enough, the electromagnetic waves reflected from it will seem to come from an image antenna that forms the missing half of the dipole. In the case of the hybrid plasmonic waveguide (as shown in Fig. 1), the metallic part of the waveguide serves as the conducting ground. The antenna feed can thus be placed at the terminal and subsequently the monopole antenna on top of it to form a nanocavity.

The maximum power that can be coupled from the waveguide into the feed depends on the design of the antenna and nanocavity. Both factors have their respective underlying design principles, but the common principle is the resonance condition. The antenna requires a resonance length which is unique to a particular optical frequency for maximum optical coupling, while the nanocavity has to form an optical resonator to allow effective localization and absorption of plasmonic power. We will discuss these two factors in detail.

The design of plasmonic antenna is analogous to that of the radio-frequency (RF) antenna, which indicates that the resonance lengths of an antenna that matches to an input electromagnetic wave are integer multiples of wavelengths: $L_r = n(\lambda/2)$ for dipoles and $L_r = n(\lambda/4)$ for monopoles [20]. However, for plasmonic antennas that operate at optical or near infrared frequencies, the resonance lengths are significantly shorter than those operating at radio frequencies. For instance, an 80nm aluminium monopole antenna is resonant to light of 514nm wavelength [21], while the first resonance length of a gold dipole is 250nm in response to light of 830nm wavelength [22].

The shortening of the resonance lengths for the infrared and optical frequencies is largely due to the resonance condition of the antenna being changed according to the finite permittivity of metals [23 – 26]. The resonance condition of the antenna depends on the effective wavelength, which depends on the permittivity of the dielectrics and the metal. For a simplified 2D metal-dielectric interface, the effective wavelength $\lambda_{eff}$ is given as:

$$\lambda_{eff} = \left(\frac{\varepsilon_s + \varepsilon_m}{\varepsilon_s \varepsilon_m}\right)^{1/2} \lambda \qquad (1)$$

where $\varepsilon_s$ is the permittivity of dielectrics and $\varepsilon_m$ the permittivity of metals. At radio frequencies, $\varepsilon_m$ is near infinite, but at visible and near infrared frequencies it drops to the order of $10 - 10^3$. Meanwhile, $\varepsilon_s$ is relatively unchanged for different frequencies. From the equation, it can be seen that when $\varepsilon_m$ is finite the effective wavelength will be smaller and depend considerably on $\varepsilon_m$.

The 2D $\lambda_{eff}$ model is insufficient to explain the optical response of a 3D metallic antenna. This is because at the high frequency regime, electromagnetic wavelength is comparable to the skin depth of the metal, resulting in the skin depth being larger than the finite-sized antenna elements [23 – 26]. Thus $\lambda_{eff}$ of the antenna at nanoscale is shape-dependent. Taking a rod-shaped metal particle as an example, $\lambda_{eff}$ is given by [25]:

$$\lambda_{eff} = \frac{\lambda}{\sqrt{\varepsilon_s}} \sqrt{\frac{x}{1+x}} - \frac{4r}{N} \qquad (2)$$

and $$x = 4\pi^2 \varepsilon_s \left(\frac{r^2}{\lambda^2}\right) \left[a_1 - a_2\left(\frac{\varepsilon_\infty + a_3\varepsilon_s}{\varepsilon_s}\right) + a_2\left(\frac{\sqrt{\varepsilon_\infty + a_3\varepsilon_s}}{\varepsilon_s}\right)\frac{\lambda}{\lambda_p}\right]^2 \qquad (3)$$

where $r$ is the rod radius, $\lambda_p$ the plasma wavelength, $N$ the order of resonance, $a_1 = 13.74$, $a_2 = 0.12$ and $a_3 = 141.04$ are numerical fit constants found in ref. [25] and $\varepsilon_\infty = \varepsilon_m(\lambda \to 0)$ the infinite frequency limit of dielectric function of metal. The first term in Eq. (2) takes into account the effect of finite radius (thickness) of the antenna on the wavelength scaling. The subtraction of the second term $4r/N$ meanwhile takes into account the apparent increase of the antenna length due to the reactance of the rod ends. Longer rod lengths can support higher order resonances (quadrupoles, sextupoles, etc.), but the higher order spectral responses are not direct integer-multiples of the first $\lambda_{eff}$. As $N$ increases, the subtraction of $4r/N$ becomes smaller and thus the higher order resonances shifts to longer $\lambda_{eff}$.

For farfield antennas, when the antenna length is matched to the electromagnetic wavelength, the resonance allows for power to be coupled to the whole conducting antenna, and eventually delivered to the feed via electrical conduction. However, in the design of waveguide-coupled nanocavity antennas based photodetector, the main objective is to deliver power directly into the nanocavity for absorption and generation of electron-hole pairs. Thus an additional design rule applies: the current distribution at the center-feed where the nanocavity is located has to be maximum. Only the odd orders (odd integer multiples) display such current distribution. For even orders, though the power reception is still maximum, the current distribution is spread along the arms of the antenna as depicted in ref. [20, pg. 19]. The plasmonic power that is distributed along the antenna will be radiated or absorbed by the metallic antenna, not contributing to the generation of electron-hole pairs.

Meanwhile, to obtain maximum power localization in the nanocavity, the nanocavity should behave as an optical resonator. A resonant cavity can be characterized by the formation of a standing wave which provides for positive optical feedback similar to microwave and laser cavities. One simple design of the nanocavity would be plane-parallel interfaces, also called the Fabry-Pérot resonator arrangement [27, 28]. The fundamental mode of this type of cavity is of half the effective wavelength. The effective wavelength is dependent on the geometrical parameters as well as the material inside the nanocavity. The cavity length should thus be constructed as integer multiples of half effective wavelengths to obtain maximum confinement of plasmonic power.

## 3. Simulation Results and Discussion

The schematic diagram of our hybrid-waveguide-coupled nanocavity with a monopole antenna design is shown in Fig. 1. The hybrid waveguide consists of a 150nm silica slot (ε = 2.085) sandwiched in between a 400nm wide metallic aluminium (ε = –252.5 + 46.07i, $\lambda_p$ = 96.7nm and $\varepsilon_\infty$ = 1) and a 150nm wide silicon (ε = 12.11).

At the end of the waveguide is a monopole antenna constructed by a metallic aluminium nanorod, directly mounted onto the aluminium slab. Sandwiched in between the nanorod and the waveguide is the antenna feed, which is filled by absorption material germanium (ε = 18.28 + 0.0485i) that forms a metal-semiconductor-metal (MSM) photodetector. This type of cavity structure is chosen for its high frequency and large sensitivity responses, ideal for terahertz (THz) applications [19]. The whole structure is planar and 220nm thick. In our simulations, we assume a 1.55μm wavelength signal is injected into the waveguide and guided by the hybrid waveguide to the nanocavity. The whole structure is simulated using finite integral time domain method in the transient solver of CST Microwave Studio [29]. Only the fundamental TM mode is considered in the simulation due to the nature of hybrid plasmonic waveguide. All material parameters are obtained from Palik's handbook [30].

Fig. 2(a) shows the effect of the antenna length on the enhancement of power absorption in the nanocavity. The first three resonance lengths are found to be 150nm, 350nm and 600nm which correspond to a maxima, a minima and another maxima, respectively. The resonance lengths are found to be much shorter than quarter-multiples of free space wavelengths due to the finite permittivity of aluminium and shape-dependence of the antennas at 1.55μm wavelength.

First, from a 2D model outlook, the first resonance length should be roughly obtained as 267.3nm from Eq. (1) (assuming that $\varepsilon_s$ is predominantly silica). This is just slightly smaller than the resonance length obtained according to classical antenna theory where it should be 268.4nm. However the new value is still very much larger than 150nm. This is because the antenna at nanoscale is shape-dependent due to the skin depth of aluminium comparable to the thickness of the antenna.

From a 3D model outlook from Eq. (2), we can see that the finite radius of the antenna plays an important role on the effective wavelength. The antenna radius in our simulation is approximately ~100nm, which renders the term $x \gg 1$ in Eq. (3) and thus the square-root term $\sqrt{\dfrac{x}{1+x}}$ in Eq. (2) can be approximated to unity. Hence a good estimation of the shape effect can be done by simply subtracting the rod-end reactance term $4r/N$ found in Eq. (2) from the 2D $\lambda_{eff}$. For the first resonance length the subtraction of the reactance term yields a resonance length of 167.3nm, very close to the simulation value of 150nm.

From Fig. 2(a), it can also be seen that higher order resonance lengths are not direct integer-multiples of the first resonance length due to the shape effect mentioned previously. From the simulation results, the first resonance length $L_{fr}$ is found to be 150nm, while the second order resonance length is 350nm, longer than 2$L_{fr}$ and the third order resonance is 600nm, longer than 3$L_{fr}$. This is because the reactance term $4r/N$ of the higher order resonances is smaller, thus subtraction of a smaller term shifts the higher order modes to longer resonance lengths.

Also, it should be noted that the antenna system in the simulation is complicated as the antenna interfaces with several different materials, such as germanium, silicon and silica. The effective background dielectric constant $\varepsilon_s$ is thus not consistent in each segment of the antenna, and it would have an effect on the resonance condition as the antenna length changes. For example, the 150nm antenna interfaces with the germanium nanocavity, silica and silicon waveguides at roughly proportional area sizes, but for the 600nm antenna the interfacial dielectric material is predominantly silica.

The inset of Fig. 2(a) depicts the current distribution profiles of the first three resonances. It is seen in Fig. 2(a, ii) that the current distribution of an even order resonance is minimum at the center feed. Meanwhile, Fig. 2(b) shows the current distribution intensity map from the simulation results. There is a noticeable current redistribution from the feed to the arm of the 350nm antenna (indicated by a white arrow), which is characteristic to the second order current density profile. Thus for waveguide-coupled nanocavity antennas, only odd order resonances should be used. At odd resonance lengths, there is a 12 times increase in power absorption compared to that without using an antenna.

The localization and absorption of plasmonic power depend strongly on the dimensions of the nanocavity. Fig. 3(a) shows how the power absorption changes with cavity length, with the two peaks correspond to the first longitudinal cavity mode seen in Fig. 3(c) and the second longitudinal cavity mode seen in Fig. 3(d) respectively. Fig. 3(c) represents the first order longitudinal cavity mode with cavity length of 150nm, and it is characterized by antinodes held at two ends of the cavity interface and a node at the center, forming a half-wavelength standing wave. Fig. 3(d) meanwhile represents the second order longitudinal cavity mode obtained by extension of the cavity length to 300nm, characterized by a full-wavelength standing wave.

It is also observed from Fig. 3(a) that the power absorption for the second longitudinal mode cavity length at 300nm is higher than that for the first longitudinal mode cavity length at 150nm, when both structures are optimized to their respective resonant antenna lengths. This is because the 300nm cavity length has a larger and longer active absorption volume, thus providing for higher absorption of plasmonic power. However, it should also be reminded that longer nanocavities will reduce device compactness, which is not the motivation for the development of plasmonic devices.

Ideally the cavity width should be small to obtain high electric-field intensities. It is shown before that the electric-field intensity, and hence instantaneous power, increases with decreasing cavity width [13]. The principle behind this phenomenon is described as analogous to that in a parallel-plate capacitor, where a smaller distance between the plates leads to higher capacitance and hence larger electric-field intensities inside the gap. However, this larger electric-field intensity resulting from a smaller cavity width does not translate to higher power absorption due to a smaller volume for light-matter interaction. Thus there is an optimal cavity width for a nanocavity to localize and absorb the coupled electromangetic waves. This has been demonstrated in Fig. 3(b), where the highest power absorption is observed at a cavity width of 60nm optimized to a resonant cavity length. While for other widths, the power does not go as high even though the cavity is resonant.

The best performance of our nanocavity with a monopole antenna structure is shown at a resonant nanocavity dimension of 150nm × 60nm, and when coupled with a resonant antenna, it translates to 42% in power absorption. In comparison to the dipole antenna design [13] scaled and optimized to similar resonant nanocavity and antenna dimensions, the power absorption only goes as high as 27%. The proposed monopole antenna based nanocavity hence improves the absorption by 55%. The lower power absorption in the dipole antenna based nanocavity design is partly due to the 'leaky waves' that travels along the metal-insulator-metal waveguide formed by the metallic slab and the antenna arm indicated by the white arrow in Fig. 4(b). This waveguide "gap" is non-existent in the monopole design in Fig. 4(a); instead the plasmonic waves are reflected from the ground interface into the nanocavity, thus enhancing the power concentration in the nanocavity.

## 4. Considerations in design of plasmonic photodetectors

To apply the proposed nanocavity for designing a plasmonic photodetector, two electrodes are needed to apply a bias voltage and conduct the photocurrents generated. The antenna and the aluminum slab are conducting and in contact with the nanocavity. The challenge remains how to extend the electrodes from there to the external circuitry.

For the electrode attached to the antenna, it should be perpendicular to the antenna arm to avoid affecting the resonance conditions. While for the other electrode that connects the aluminum slab, it should not be in parallel with the former to avoid forming another MIM waveguide which may guide and draw power out from the nanocavity. As such, the electrode can extend out from the length side of the aluminium slab in direction perpendicular with the antenna-electrode. One suggested placement of electrodes is shown in the inset in Fig. 5, where two aluminium wires (cross section 150nm×220nm) are employed. Simulation results show that this arrangement of electrodes does not compromise on the device integrity, bearing only less than 2% loss in power absorption. This is in contrast to the dipole where electrodes need to be made out of transparent conducting oxide (TCO) material to avoid possible effects on the antenna resonance [14]. The use of TCO material for electrodes is generally not encouraged due to its higher resistance compared to metal electrodes.

In the aspect of device fabrication, the monopole antenna design is easier to fabricate than the dipole antenna, owing to the gapless feature of the former. In its monolithic build, there will not be any alignment issues because the nanocavity and antenna unit is not isolated from the waveguide, allowing for inherent self-alignment. There is also the advantage of simpler lithographic mask sets because of the absence of the gap and one of the arms of the antenna. Also because the electrodes can be made out of the same material with the antenna and metallic aluminium slab, one fabrication step which involves the deposition of TCO electrodes is reduced, saving the cost of introducing extra materials in the fabrication of the device. Moreover, the materials chosen in our discussion are CMOS compatible, which is an advantage to realize it with conventional CMOS technology.

## 5. Conclusions

We have demonstrated a waveguide-coupled nanocavity with a monopole antenna to localize and absorb the optical power from a hybrid plasmonic waveguide. The monopole antenna is directly attached at the hybrid waveguide terminal, with the nanocavity sandwiched in between the antenna and the metallic part of the waveguide. Maximum power absorption in the nanocavity can be achieved by carefully designing the structure with a resonant antenna and resonant nanocavity. The design of the resonant antenna follows the frequency-dependent conductivity of the metal and also the shape of the antenna, and only odd order resonances are used to ensure current distribution is maximum in the nanocavity. Meanwhile, the nanocavity is designed as a Fabry-Pérot resonator for effective localization of the coupled plasmonic waves. There is an optimal cavity width and length for the resonant nanocavity to localize and absorb maximum optical power. Simulation results show that 42% of optical power from the waveguide can be absorbed in a germanium filled nanocavity with a volume of $220\times150\times60nm^3$, when the length of the antenna is 600nm, and the width and length of the nanocavity are 60nm and 150nm repectively. In comparison to the nanocavity formed with a dipole antenna, the proposed monopole antenna based nanocavity shows 55% improvement in optical power absorption. In addition, the proposed nanocavity has several advantages over the dipole antenna based nanocavity in design of plasmonic detectors. Metal instead of TCO materials can be used for the electrodes, and the device structure is inherently self-aligned, leading to a high-performance, ultra-compact and easy-to-fabricate plasmonic detectors. This proposed nanocavity design can be used in many aspects of plasmonic nanocircuits, including but not limited to plasmonic photodetectors, switches, waveguide modulators, waveguide emitters and sensors. This design would bring us one step closer to integration of high-speed optical and compact electronic devices into the same chip to fully take the advantages of optics and nanoelectronics.

## 6. Acknowledgement

This research is supported by A*STAR SERC Grant, Metamaterials Programme: Nanoplasmonics. Grant No.: 092 154 0098. One of us (KJAO) is supported by a PhD scholarship funded by the MOE Tier2 grant (2008-T2-01-033).

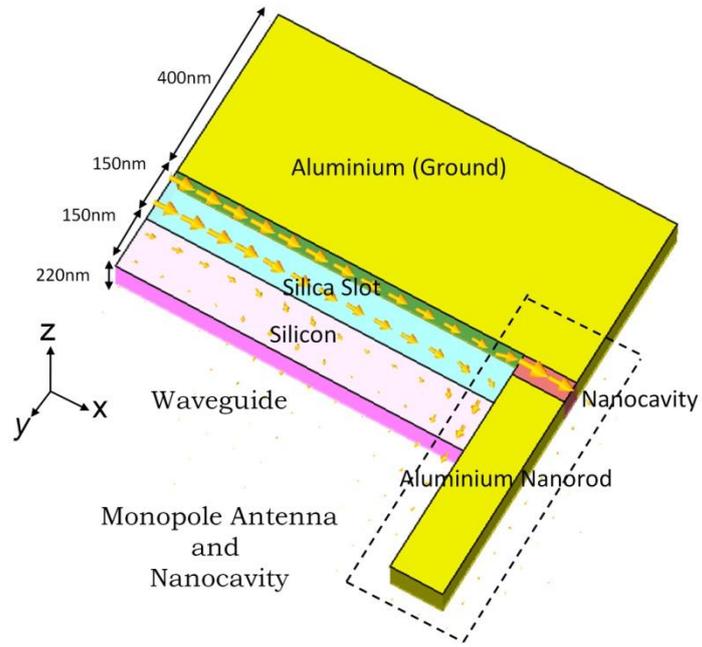

Fig. 1. (Color online) Schematic diagram of a plasmonic hybrid waveguide coupled nanocavity with a monopole antenna. The antenna is directly mounted onto the aluminium slab, and a nanocavity is formed at the antenna feed. The plasmonic waves, represented by arrows, are guided by the hybrid waveguide and focused into the nanocavity.

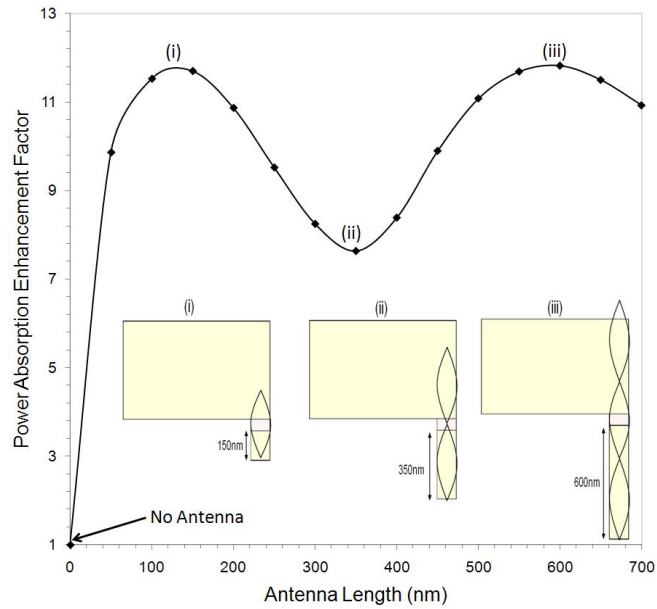

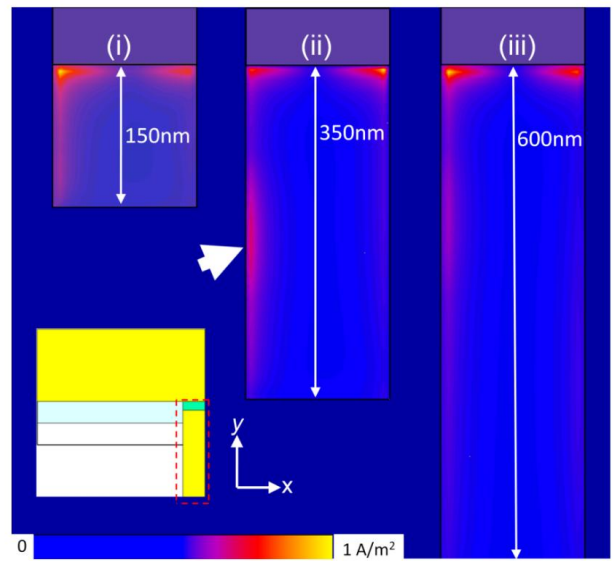

Fig. 2. (Color online) (a) Power absorption enhancement factor as a function of antenna length. The first three resonant lengths found are 150nm, 350nm and 600nm. The higher order resonance lengths are not direct integer-multiples of the first resonance length due to the shape effect. Inset: 1D current density profiles for antenna nanorod lengths of (i) 150nm, (ii) 350nm and (iii) 600nm. (b) Current distribution maps for resonant antenna lengths of (i) 150nm, (ii) 350nm and (iii) 600nm. The portion plotted is circled in the inset. There is a noticeable current redistribution from the base to the length side of the 350nm antenna (indicated by a white arrow), which is characteristic to the second order current density profile.

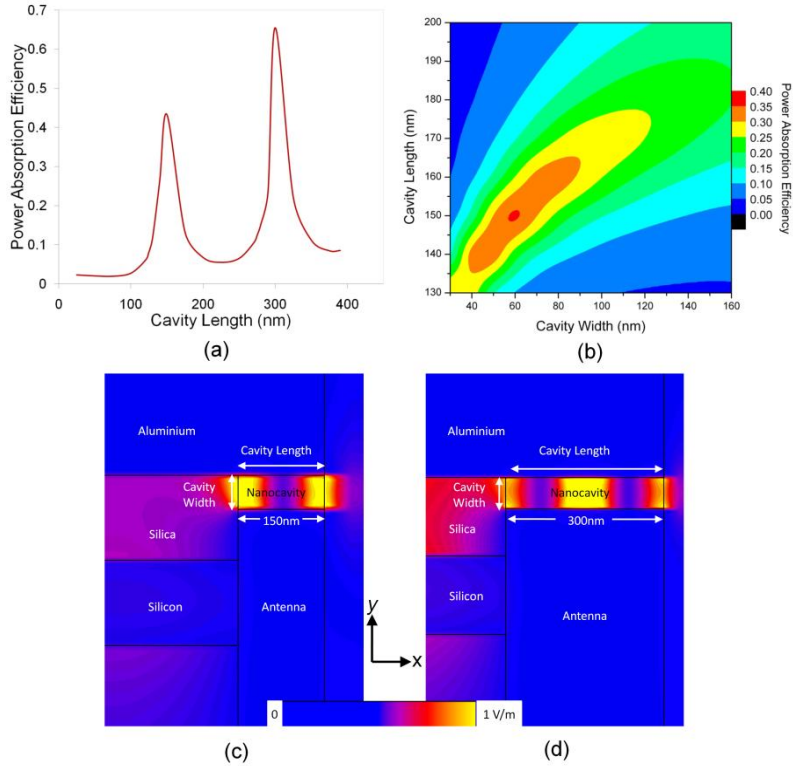

Fig. 3. (Color online) (a) Power absorption efficiency as a function of cavity length for a nanocavity with 60nm cavity width and 600nm resonant antenna. The two peaks correspond to the first and second longitudinal cavity modes respectively. (b) Power absorption efficiency contour plot for nanocavities with a 600nm resonant antenna. The best performing resonant nanocavity dimension is found to be 150nm x 60nm, with recorded power absorption of 42%. (c) and (d) Electric field intensity maps of (c) a 150nm half-wavelength nanocavity and (d) a 300nm full-wavelength nanocavity.

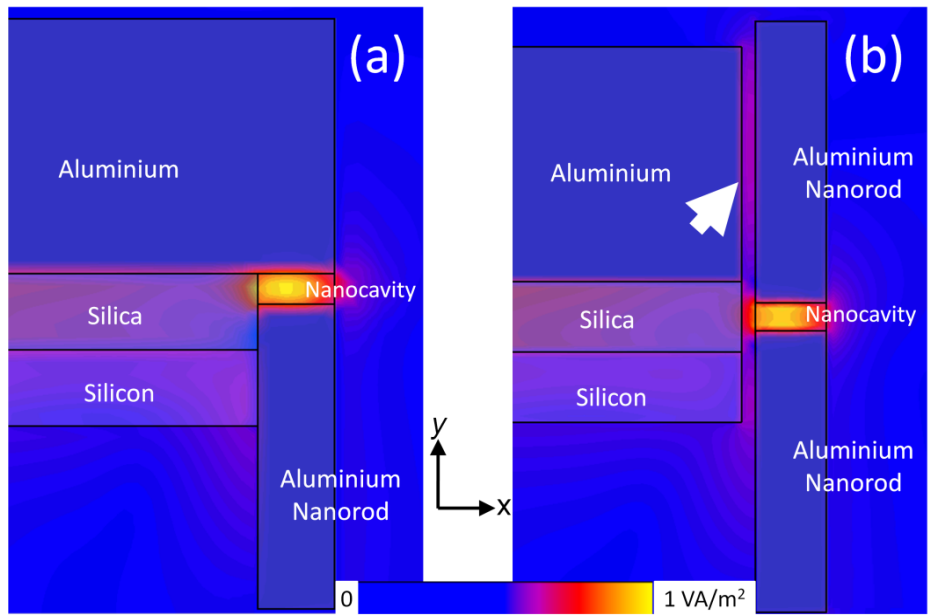

Fig. 4. (Color online) Power flow intensity maps of a hybrid-waveguide-coupled nanocavity with (a) monopole antenna and (b) dipole antenna. The power coupling from the waveguide to the dipole antenna nanocavity is lowered due to 'leaky waves' travelling along the metal-insulator-metal waveguide that is formed by the antenna arm and the aluminium slab (indicated by a white arrow).

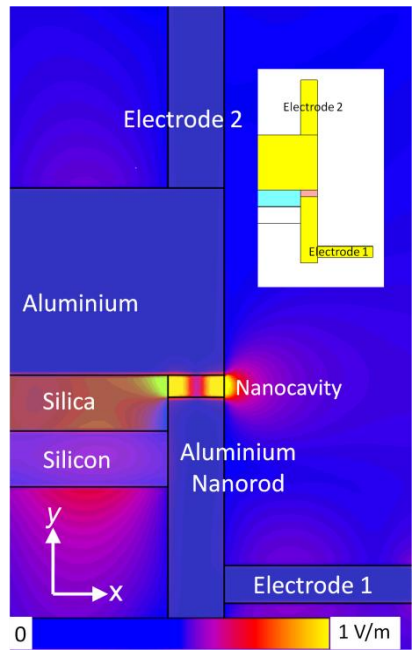

Fig. 5. (Color online) Electric field intensity map of the plasmonic photodetector with attached electrodes. The electrode attached to the antenna (electrode 1) is placed in perpendicular to the antenna arm, while the electrode that connects the metallic ground (electrode 2) is extended out from the aluminium slab perpendicular in direction with the antenna-electrode (inset).